\documentclass[twocolumn, showpacs,preprintnumbers]{revtex4}
\usepackage{amsmath,amssymb}
\usepackage{graphicx}
\usepackage{dcolumn}
\usepackage{bm}

\usepackage{color}

\def \dd#1{\null}

\begin{document}

\title{Deterministic source of a train of
indistinguishable single-photon pulses with single-atom-cavity system}
\author{A.Gogyan,$^{1,2}$ S. Gu\'erin,$^{2}$ H.-R. Jauslin,$^{2}$ and Yu. Malakyan$^{1,3}$}
\email{yumal@ipr.sci.am}
\affiliation{$^{1}$ Institute for Physical Research, Armenian National Academy of Sciences, Ashtarak-2,
0203, Armenia}
\affiliation{$^{2}$Laboratoire Interdisciplinaire Carnot de Bourgogne, UMR CNRS 5209, BP 47870, 21078 Dijon, France}
\affiliation{$^{3}$ Centre of Strong Field Physics, Yerevan State University, 1 A. Manukian St., Yerevan 0025, Armenia}


\date{\today }

\begin{abstract}
We present a mechanism to produce  indistinguishable single-photon pulses on demand from an optical cavity. The sequences of two
laser pulses generate, at the two Raman transitions of a four-level atom, the same cavity-mode photons without repumping of the
atom between photon generations. Photons are emitted from the cavity with near-unit efficiency in well-defined temporal modes
of identical shapes controlled by the laser fields. The second order correlation function reveals the single-photon nature of
the proposed source. A realistic setup for the experimental implementation is presented.
\end{abstract}

 \maketitle




Deterministic sources of high-quality single-photon (SP) states
are of great importance for quantum information
processing \cite{monroe}. A basic requirement for many quantum
optics applications, including quantum computing with linear
optics \cite{knill,dowling}, quantum cryptography \cite{gisin} and
entanglement swapping \cite{pan}, is to have single photon
pulses with well-defined identical shapes, frequency and
polarization, as these  schemes  based
on photon-interference effects are very sensitive to the
parameters of SP pulses and their repetition rate. A good source has to ensure a pure SP
state without mixture from both the multi-photon and zero-photon
states, as well as to prevent the entanglement
between the photons which degrades the purity of the SP state.
Since the individual photons are usually emitted during the
spontaneous decay of atomic systems, the SP sources must be immune
to the environmental effects that induce the dephasing of atomic
transitions. Most of the schemes proposed earlier to produce
single photons on demand from  solid state single emitters
\cite{sanders}, organic molecules \cite{lounis,treussart}, and
quantum dots \cite{michler, santori} are confronted with this
difficulty. Besides, they do not offer a high efficiency because
of the isotropic nature of fluorescence that prevents to collect
the photons, not to mention the spectral dephasing and
inhomogeneity of solid-state emitters. Deterministic sources of single photons are realized also in cold atomic ensembles with feedback circuit \cite{Matsukevich, Chen}. But these schemes are not suitable to generate SP train with an arbitrary repetition rate because of strong temporary bounds caused by the feedback and write-read processes.

At present, all the requirements mentioned above can be  achieved together
with a $\Lambda$-type atom trapped in high-finesse optical
cavities \cite {kuhn,mckeever,keller,hijlkema}, where the single
photons are generated via vacuum-stimulated Raman scattering of a
classical laser field into a cavity mode.  These systems not
only provide a strong interaction between a photon and an atom,
but also support very high collection efficiency due to the fact
that the photons leave the cavity through one mirror with a
transmissivity incomparably larger than that of the opposite one.
By carefully adjusting the parameters of the laser pulse one can
also easily control the waveform of output single photons.
However, the main disadvantage of these schemes is the necessity
to use a repumping field to transfer the population of the atom to
its initial state after the generation of a cavity photon and
only then to generate the next one. In this Letter we propose a scheme featuring a double Raman atomic
configuration, which is able to deterministically generate a
stream of identical SP pulses without using the repumping field,
while maintaining the high generation efficiency, as well as providing simpler control of the output
photon waveforms. It is interesting to note that a repumping field
is not required also in a similar scheme for generating a sequence
of single photons of alternating polarization \cite{wilk}. 
\begin{figure}[b] \rotatebox{0}{\includegraphics*
[scale =0.7]{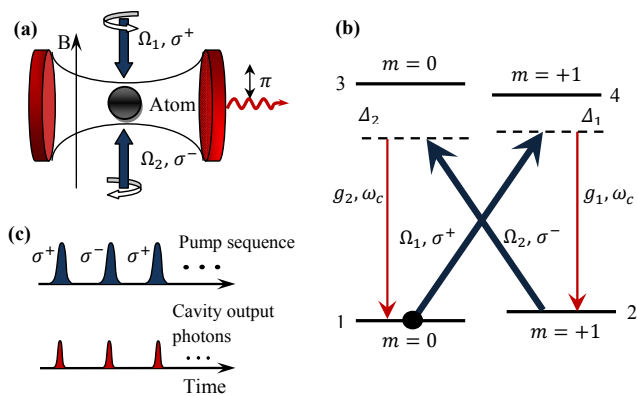}} \caption{(Color online) Schematic setup. (a) A
single atom trapped in a high-Q cavity is driven by two laser
pulses. b) Relevant atomic level structure in an external magnetic
field. The case is shown, when the Land\'e $g_L$-factors of
the ground and excited states have opposite signs.  c) Sequence of
laser pulses and generated cavity output single-photons.}
\end{figure}

Our scheme, illustrated in Fig. 1, involves a four-level atom
trapped in a one-mode high-finesse optical cavity. The two ground
states 1 and 2 and two upper states 3 and 4 of the atom are Zeeman
sublevels (Fig. 1b), which are split by a magnetic field acting
perpendicular to the cavity axis. The atom is initially prepared in one of
the ground states, for instance, in state 1 of magnetic quantum
number $m=0$, and interacts in turns with a sequence
of two pumping fields as shown in Fig. 1a. At first, a coherent
$\sigma^+$ polarized field of Rabi
frequency $\Omega_1$ applied between ground state 1 and excited
state 4 transfers the atom to ground state 2 while creating a
cavity-mode Stokes-photon. Then, after a programmable delay time
$\tau_d$, the $\sigma^-$ polarized pump pulse $\Omega_2$ generates
the anti-Stokes photon at the $3\rightarrow 1$ transition and
transfers the atom back to  ground state 1. The Stokes and
anti-Stokes photons have identical frequencies, so that the cavity
is coherently coupled to the atom on both transitions
$4\rightarrow 2$ and $3\rightarrow 1$ with the rates $g_1$ and
$g_2$, respectively, resulting in the generation of linearly
polarized cavity-photons in both cases. The laser fields are tuned
to the two-photon resonance, while the one-photon detunings are
very large compared to the Rabi frequencies and the cavity damping
rate $k$: $\Delta_i\gg g_i, k, \Omega_i,i=1,2$. This condition
makes the system robust against the spontaneous loss from upper
levels and dephasing effects induced by other excited states. More
importantly, in the off-resonant case the Raman process with
effective atom-photon coupling $G_i=g_i\Omega_i/\Delta_i,
i=1,2,$ can be made much slower than the cavity
field decay: $G_i\ll k$. This ensures that a generated photon
leaves the cavity long before the next cavity-photon is emitted
and, hence, no entanglement between the photons will be created,
if we also take into account that the coherence between atomic
ground states is always zero. Therefore, we can construct identical wavepackets for outgoing photons independently from
each other, as they are entirely determined by the temporal shape
of the corresponding pump pulse.
 A remarkable feature of our scheme is that, despite the smallness
 of $G_{1,2}$, it is able to produce cavity photons with near-unit
 efficiency, as discussed in more detail below.

We first solve the equations for the number of cavity photons and
for their flux giving the SP detection time distribution, and
calculate the correlation between the output photons. We carry out
numerical calculations for realistic atomic systems in two cases.
The first one, shown in Fig. 1, is $^{85}Rb$ atom with the ground
states $5S_{1/2}(F=2, m_F=0)$, $5S_{1/2}(F=2, m_F=+1)$ and exited
states $5P_{1/2}(F^{\prime}=1, m_{F^{\prime}}=0)$,
$5P_{1/2}(F^{\prime}=1, m_{F^{\prime}}=+1)$  as state 1,2 and 3,4
in our scheme, respectively. The central drawback of this scheme
is that the spontaneous decay of state $5P_{1/2}(F^{\prime}=1,
m_{F^{\prime}}=+1)$ into the ground state $5S_{1/2}(F=1)$
(level not shown) constitutes a loss channel that moves the system
outside the considered level configuration. However, we show that
even in this case the atom can generate about 80 identical SP
pulses before falling into the ground $F=1$ state. To restore the
generation a repumping field must be applied to transfer the atom
into the initial state. For continuous generation of SP, we
consider a second case employing cycling transitions of the
$D_2$ line in the  $^{85}Rb$ atom with $5S_{1/2}(F=2)$ and
$5P_{3/2}(F^{\prime}=3)$ as the ground and excited states. The
main limitation in this case is the atom life-time in
cavities, which amounts to at most one minute
\cite{hijlkema}. In both cases, to analyze the dynamics of a
coupled atom-cavity system under realistic conditions, we need to
determine how other Zeeman sublevels, not shown in Fig. 1, impact
the generation process. It is easy to see, however, that it does
not matter what Zeeman  sublevel of the ground state is initially
populated. Indeed, in any case the atom emits cavity photons many
times before jumping into a new pair of ground Zeeman sublevels
via spontaneous decay of upper states; the generation of cavity
photon from this new pair of Zeeman sublevels occurs in the above
described manner, provided that $\Delta_i \gg \Delta_B$,
where $\Delta_B=g_L\mu_B B$ is Zeeman splitting of atomic levels
in the magnetic field $B$ with $g_L$ the Land\'e factor and
$\mu_B$ Bohr magneton.

The laser fields propagating perpendicular to the cavity axis are given by
\begin{equation}
E_j(t)={\cal E}_jf_j^{1/2}(t)\exp (-i\omega _jt),\ j=1,2.
\end{equation}
where $f_1(t)=\underset{l=1}{\overset{N}{\sum }} f_1^l(t)$
and $f_2(t)=\underset{l=1}{\overset{N}{\sum }}
f_2^l(t-\tau_d)$ represent the sum of $N$ well-separated temporal
modes with profiles $f_1^l(t)$ and $f_2^l(t-\tau_d)$ for the
$l^{th}$ mode in the pump series 1 and 2, respectively. ${\cal
E}_j$ is the peak amplitude of the field $j.$

In far off-resonant case, we can adiabatically eliminate the upper atomic states 3 and 4 and write the effective Hamiltonian as
\begin{equation}
H=\hbar [ G_1f_1^{1/2}(t)\sigma
_{21}+G_2f_2^{1/2}(t)\sigma _{12}]a^{\dag} +h.c.
\end{equation}
with $\sigma_{ij}$ and $a(a^\dag)$ the atomic and cavity mode operators, respectively. The peak Rabi frequencies of the laser fields are
given by $\Omega _{1}=\mu_{41}E_{1}/\hbar, \Omega _{2}=\mu _{32}E_{2}/\hbar$  with $\mu _{ij}$ the dipole matrix element of the $i\rightarrow j$
transition.

For $\Omega_i$ and $g_i$ of the same order, the Stark shifts of the atomic ground levels, of the form
$\Omega_i^2f_i(t)/\Delta_i$ and $g_i^2/\Delta_i$, are negligibly small with respect to the cavity linewidth $k$ in the bad cavity
limit $G_i\ll k$ as considered here.

The system evolution is described by the master equation for the whole
density matrix $\rho $\ for the atom and cavity mode \cite{gard}
\begin{equation}
\ \frac{d\rho }{dt}=-\frac{i}{\hbar }[H,\rho ]+\frac{d\rho }{dt}|_{rel}
\end{equation}
where the second term in the right hand side (rhs) accounts for all relaxations
in the system. With the use of the Lindblad operator $L[\hat{O}]\rho =\hat{O}
\rho \hat{O}^{\dag}-(\hat{O}^{\dag}\hat{O}\rho +\rho \hat{O}^{\dag}\hat{O})/2$ it is
written\ in the form
\begin{equation*}
\frac{d\rho }{dt}|_{rel}=kL[a]\rho +\Gamma _1(t)L[\sigma _{21}]\rho
\end{equation*}
\begin{equation}
\quad \quad +\Gamma _{2}(t)L[\sigma _{12}]\rho -(\Gamma
_{1,\text{out}}\sigma_{11}+\Gamma _{2,\text{out}}\sigma_{22})\rho
\end{equation}
The first term in the rhs of this equation represents the cavity output coupling, while the second and third terms
describe the optical pumping to ground states 2 and 1 from the states 1 and 2, respectively, and give rise to noise
of corresponding rates
\begin{equation}
\Gamma _{1}(t)=\frac{\Omega _{1}^{2}}{\Delta^{2}}f_{1}(t)\gamma _{42},\
\ \ \Gamma _{2}(t)=\frac{\Omega _{2}^{2}}{\Delta^{2}}f_{2}(t)\gamma
_{31}
\end{equation}
The last term of Eq.(4) introduces the losses of atomic population due to the decay of upper atomic states 3 and 4
 into states outside of the system with rates $\gamma_{3,\text{out}}$ and $\gamma_{4,\text{out}}$, respectively:
\begin{subequations}
\begin{eqnarray}
\Gamma_{1,\text{out}}(t)=\frac{\Omega^2_1}{\Delta_1^2}f_1(t)\gamma_{4,\text{out}} =\Gamma_{1,\text{out}} f_1(t), \\
\ \Gamma _{2,\text{out}}(t)
=\frac{\Omega^2_2}{\Delta_2^2}f_2(t)\gamma_{3,\text{out}}=\Gamma_{2,\text{out}} f_2(t).
\end{eqnarray}
\end{subequations}
Spontaneous emission channels corresponding to (i) the cycling
transition returning the atom back to the starting state and to
(ii) the atomic transfer to other ground sublevels with $\Delta
m_F=\pm 2$, that lead to emitted photons not in the cavity mode,
should be in principle included. However, in contrast to the
decays of Eqs. (5) and (6), these channels neither change the
population of the system, nor lead to any noises, therefore, they
can be ignored.

Our aim is to find the flux of the output photons
\begin{equation}
dn_{out}(t)/dt = \langle a^{\dag}_{out}(t) a_{out}(t)\rangle
\end{equation}
which describes the sequence of outgoing SP wavepackets. Here
$n_{out}(t)$ is the mean photon number of the output field from
the cavity $a_{out}(t)$, which is connected with the input
$a_{in}(t)$ and intracavity $a(t)$ fields by the input-output
formulation \cite{gard}
\begin{equation}
a_{out}(t)+a_{in}(t)=\sqrt{k }a(t)
\end{equation}
and satisfies the commutation relation $[a_{out}(t),a_{out}^{\dag}(t^{\prime })]=[a_{in}(t),a_{in}^{\dag}(t^{\prime})]=\delta (t-t^{\prime })$.
With the Hamiltonian (1), the Heisenberg-Langevin equation for $a(t)$ is given by \cite{gard}
\begin{equation}
\dot{a}=-iG_1f_1^{1/2}(t)\sigma_{21}-iG_2f_2^{1/2}(t)\sigma_{12}-(k/2) a+\sqrt{k}a_{in}(t)
\end{equation}
In the bad-cavity limit $k\gg G_{1,2}$, we adiabatically eliminate the cavity mode $a(t)$ yielding
\begin{equation}
 a=-\frac{2i}{k}[G_1f_1^{1/2}(t)\sigma_{21}+G_2f_2^{1/2}(t)\sigma_{12}]+\frac{2}{\sqrt{k}}a_{in}(t)
\end{equation}

Upon substituting this solution into the Hamiltonian,  for the
case of a vacuum input $\langle a^\dag_{in}(t)a_{in}(t)
\rangle=0$, we eventually obtain from Eq.(3) the following
equations for the atomic variables $i,j=1,2; j\neq i$
\begin{figure}[b] \rotatebox{0}{\includegraphics*
[scale =0.9]{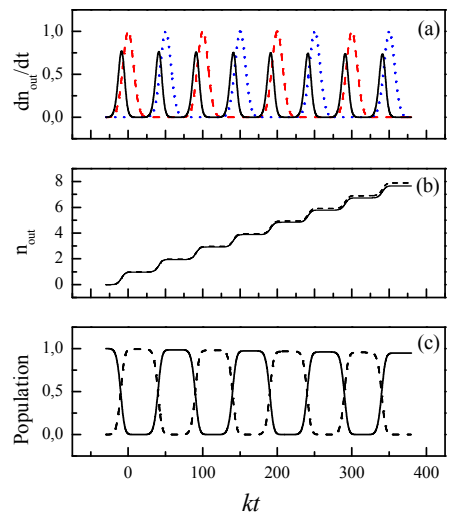}} \caption{(Color online) Flux (a), total number of output field photons (b) and population of atomic ground
states 1 (solid) and 2 (dashed)(c) as a function of  time (in units of inverse cavity decay $k^{-1}$), in the case, when 4 pulses
of each laser beam are applied with delay time $\tau_d=3\mu s$. For the rest of parameters see the text.}
\end{figure}
\begin{subequations}
\begin{eqnarray}
\langle \dot{\sigma}_{ii} (t) \rangle &=&
-[\alpha_i(t)+\Gamma_i(t)+\Gamma_{i,\text{out}}(t)]\langle
 \sigma_{ii}(t) \rangle  \nonumber \\
&&+[\alpha_j(t) + \Gamma_j(t)]
\langle \sigma_{jj}(t)\rangle \\
\langle \dot{\sigma}_{21} (t) \rangle &=& - \frac{1}{2}
\underset{i=1} {\overset{2}{\sum
}}[\alpha_i(t)+\Gamma_i(t)]\langle \sigma_{21}(t)\rangle
\end{eqnarray}
\end{subequations}
which are subjected to initial conditions $\langle \sigma_{11} (-\infty)\rangle =1,
\langle \sigma_{22}(-\infty) \rangle = \langle \sigma_{21}(-\infty)\rangle=0$. Here $\alpha_i(t) = 4 G^2_i f_i(t)/k = \alpha_i f_i(t), i=1,2$.

Thus, in the bad-cavity limit the problem is reduced to the
solution of the dynamical equations for the atom. Equations
(11) are easily solved analytically. However, the final solutions
are lengthy and will be given here only graphically. We first
discuss some properties of Eqs. (11). It is seen that state 1 (and
similarly state 2) is populated in two ways: (i) via cavity photon
generation with the rate $\alpha_{1(2)}(t)$ and (ii) by optical
pumping of rate $\Gamma_{1(2)}(t)$, that gives for the
signal-to-noise ratio $R_{sn}=4g^2_{1(2)}/(k\gamma_{42(31)})$,
which must be quite large: $R_{sn}\gg1$. The second observation is
that the overall population of the atom after a total of $n$
pulses of two laser sequences for $\alpha_i T \gg 1$ decreases as
\begin{equation}
\langle \sigma_{11}(t)+ \sigma_{22} (t) \rangle  = (1 - n
\Gamma_{out}/\alpha)
\end{equation}
Here it is assumed that $\Gamma_{1out}=\Gamma_{2out}=\Gamma_{out},
\alpha_1=\alpha_2=\alpha$. Thus, the population leakage is
negligibly small until $n \alpha/\Gamma_{out} < 1$. Further, the
ground state coherence is always zero: ${\sigma}_{21}(t)=0$, as
expected due to the spontaneous nature of Raman transitions.
Finally, from the explicit expression of the flux calculated from
Eq.(7) for one pump pulse and considering $\Gamma_{out}=0$
\begin{equation}
dn_{out}(t)/dt= \alpha_1(t) e^{-\int_{-\infty}^{t} \alpha_1(t') dt'}
\end{equation}
we conclude that the waveform of the emitted single-photon is simply related to the shape of the pump pulse and, thereby, is much
easier controlled as compared to schemes proposed so far in literature. From this equation one finds
$n_{out}(t)= 1-\exp \bigl (-\int_{-\infty}^{t} \alpha_1(t') dt' \bigr )$ showing that our system is able to
 produce photons with near-unit efficiency, if $\alpha_i(t)T\gg 1$.

In Fig. 2 the calculated flux and total number of output photons,
as well as the populations of atomic ground states are shown for
the case, when each laser sequence contains four Gaussian-shaped
subpulses with duration $T=1\mu s$. We present the results of
cavity photon generation on the  D1 line transition
$5S_{1/2}(F=2)\rightarrow 5P_{1/2}(F^{\prime}=1)$  in the $^{85}Rb$
atom obtained with the following parameters:
$\Omega_{1,2}=2\pi\times 10MHz=0.1\Delta (\Delta_1 \simeq
\Delta_2=\Delta)$, and $(g_{1,2}, k, \gamma_{sp}
)/2\pi=(10,3,6)$MHz, $\gamma_{sp}$ being the total spontaneous
decay rate of the upper states. A magnetic field of 20G produces the Zeeman splitting $\Delta_B /2\pi$ =
14MHz. These parameters are within experimental reach and ensure
the fulfillment of all necessary conditions indicated above.  The
results demonstrate two important features of the scheme. The
photons are generated deterministically at the leading edge of
each pump pulse with identical duration $T_{cav}\sim T/2$ and
time-symmetric wavepackets. The  efficiency of one photon
generation by each pump pulse is close to $100\%$ (see Fig. 2b).
As one can see in Fig. 2(a,c), the peak values of the generated SP
pulses and of the atomic populations display a slight decrease in time
caused by population losses through the channel
$5P_{1/2}(F^{\prime}=1)\rightarrow 5S_{1/2}(F=1)$. Just because of
this fact the total number of emitted photons does not reach its
maximum value 8 (Fig. 2b, solid line). Nevertheless, from Eq. (12)
it follows that about $n_{out} \sim \alpha/\Gamma_{out} \simeq
R_{sn}\simeq 70$ cavity photons are generated before the losses
become significant. For comparison, $n_{out}$ is  shown also for
the lossless case of D2 line cycling transition
$5S_{1/2}(F=2)\rightarrow 5P_{3/2}(F^{\prime}=3)$ (Fig. 2b, dashed
line).
\begin{figure}[b] \rotatebox{0}{\includegraphics*
[scale =0.6]{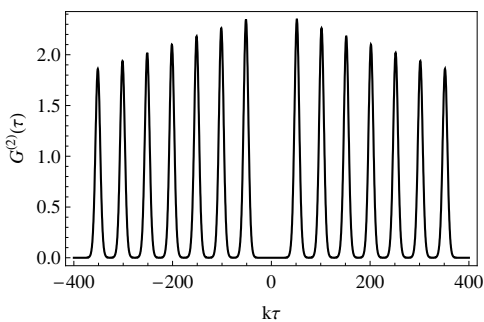}} \caption{Intensity correlation integrated over the single-photon train as a function of time delay $\tau$ between
the two photon detections. The parameters are the same as in Fig. 2}
\end{figure}

The probability of a joint detection of two photons produced in the train is given by the intensity correlation function
\begin{equation}
G^{(2)}(t,\tau)= \langle a^\dag_{out}(t)a^\dag_{out} (t+\tau) a_{out} (t+\tau)a_{out}(t)\rangle
\end{equation}
where $\tau$ is the time delay between the two photon detections. By applying the quantum
regression theorem \cite{gard,lax} and using the input-output relation Eq.(7), the second-order temporal correlation
function $G^{(2)}(t,\tau)$ is reduced to
\begin{equation}
G^{(2)}(t,\tau)= k[\alpha_1(t+\tau)Z_1(t,\tau)+ \alpha_2(t+\tau)Z_2(t,\tau)]
\end{equation}
where $Z_i(t,\tau)=\langle a^\dag(t)\sigma_{ii}(t+\tau)a(t) \rangle , i=1,2$, as a function of $\tau$, obey equations similar to Eqs.(10,11) with initial values $Z_1(t,0) = \alpha_2(t) \langle \sigma_{22}(t)\rangle /k$ and $Z_2(t,0) =
\alpha_1(t) \langle \sigma_{11}(t)\rangle/k$. Since we are interested in the total probability of a joint detection as a function of
the time delay $\tau$, we have to integrate Eq.(16) over $t$. The results of numerical calculations for
$G^{(2)}(\tau)=\int_{-\infty}^{\infty} G^{(2)}(t,\tau) dt$  are shown in Fig. 3. The temporal structure of $G^{2)}(\tau)$ reveals
the  characteristics of a pulsed source of light: the absence of a peak at delay time $\tau=0$ is evidence of the single-photon nature
of the source, and the individual peaks are separated by the pump pulses' delay. The decrease in the peak amplitude of the
probability of joint detection for increasing delay time results from having a finite train of emitted photons.

In conclusion, we have proposed a robust and realistic source  of  indistinguishable single-photons with identical
frequency and polarization generated on demand in a well-defined spatio-temporal mode from a coupled double-Raman
atom-cavity system.  The high efficiency and simplicity of the scheme, free from such complications as repumping
process and environmental dephasing, makes the generation of many SP identical pulses feasible. The removal of the repumping field is a principal task, because its usage strongly restricts the process: the repetition rate of emitting photons is limited by the acting time of the repumping field. Unlike this, our mechanism allows to freely change the repetition rate of the single-photon pulses up to zero, because even in this case non-entangled single-photons are generated. One of the most important applications of this property is to generate Fock states with a programmable number of photons. Moreover, in the good-cavity limit our scheme can serve as an one-atom laser with a controllable statistics of generated photons that provides the quantitative study of the quantum-to-classical transition in our system raised with gradual change of the parameters. These questions will be addressed in the future publications.

This research has been conducted in the scope of the International Associated Laboratory IRMAS. We also acknowledge the support from the French Agence Nationale de la Recherche (projet CoMoC), and the Marie Curie Initial Training Network Grant No. CA-ITN-214962-FASTQUAST, Scientific Research Foundation of the Government of the
Republic of Armenia (Project No.96), Armenian SCS Grant No. A-07, NFSAT Grant No. ECSP-09-85, ANSEF Grant No. PS-1993. Yu.M. thanks the Universit\'{e} de Bourgogne for his stays during which a part of this work was accomplished.

\end{document}